\documentclass[letterpaper,preprintnumbers,prd,twocolumn,nofootinbib,nobibnotes,showpacs]{revtex4}
\usepackage{epsfig}
\usepackage{graphicx}%
\usepackage{dcolumn}
\usepackage{amsmath}

\makeatletter
\def\btt#1{\texttt{\@backslashchar#1}}%
\DeclareRobustCommand\bblash{\btt{\@backslashchar}}%
\makeatother

\begin{document}

\title{Phantom Cosmic Dynamics: Tracking Attractor and Cosmic Doomsday}% Force line breaks with \\

\author{Jian-gang Hao}
\author{Xin-zhou Li}\email{kychz@shtu.edu.cn}
\affiliation{Shanghai United Center for Astrophysics (SUCA),
Shanghai Normal University, 100 Guilin Road, Shanghai 200234,China
}%

\date{\today}% It is always \today, today, but you may specify any date with \date.

\begin{abstract}
We study the dynamics of phantom model and give the conditions for
potentials to admit tracking attractor, de Sitter attractor and
big rip attractor. Especially, we show that phantom models with
exponential and inverse power law potentials do not admit the
tracking attractor solution. The "tachyonic" stability of the
system at/near the attractors as well as the quantum stabiltiy
have been studied.
\end{abstract}


\maketitle

\section{Introduction}

Astronomical observations on the CMB anisotropy
\cite{bennett,Netterfield,Halverson}, the relation between
red-shift and luminosity distance of Supernovae \cite{riess,
perlmutter,tonry} as well as the galaxy distribution, clumping and
spreading \cite{SDSS} depicted that our Universe is spatially
flat, with about two-thirds of energy density resulted from dark
energy that has an equation of state $w<-1/3$ and accelerates the
expansion of the Universe. The nature of this substance is quite
unusual and there is no justification for assuming it resemble
known forms of matter or energy. Candidates for dark energy have
been widely studied and focus on a dynamically evolving scalar
field (quintessence \cite{ratra,coble, Steinhardt,Peebles}, with
$w>-1$ and phantom \cite{caldwell}, with $w<-1$) and cosmological
constant. Present observation data constrain the the range of the
equation of state of dark energy as $-1.38<w<-0.82$
\cite{melchiorri}, which indicates the possibility of dark energy
with $w<-1$, debuted as Phantom or super quintessence
\cite{caldwell}.  The realization of $w<-1$ could not be achieved
by scalar field with positive kinetic energy and thus the negative
kinetic energy is introduced although it violates some well known
energy conditions \cite{carroll}. Another important consequence of
Phantom is the Big rip \cite{bigrip} or Big smash \cite{McInnes}
phase, in which the scale factor of the Universe goes to infinity
at a finite cosmological time. The cosmological implications of
Phantom have been widely
studied\cite{sahni,schulz,maor,Frampton,Gibbons,Sami,
Feinstein,Chimento,Dabrowski,Nojiri,meng,Johri,Cline,Lu,Gonzalez-Diaz,Stefancic}
, the Phantom model with Born-Infeld type Lagrangian has been
proposed \cite{hao3} and its generalization to brane world has
been done in Ref. \cite{liu}.

One of the most important issues for dark energy models is the
fine tuning problem and a good model should limit the fine tuning
as much as possible. The dynamical attractor of the cosmological
system has been employed to make the late time behaviors of the
model insensitive to the initial condition of the field and thus
alleviates the fine tuning problem. In quintessence models, the
dynamical system has tracking attractor that makes the
quintessence evolve by tracking the equation of state of the
background cosmological fluid so as to alleviating the fine tuning
problem\cite{attractor1,attractor2}. While in phantom models with
canonical lagrangian, the kinetic energy term becomes negative,
which make the phase space of the cosmological system behaves
rather differently from those of the quintessence. The major
difference is that the tracking attractor does not exist in the
phantom system with exponential and inverse power law potentials,
which admit tracking attractor solution in quintessence models.
Accordingly, We give the condition for the phantom system to admit
tracking attractor. On the other hand, there are also two late
time attractors in the phantom system corresponding to the big rip
phase\cite{hao1} and de Sitter phase\cite{hao2}. One cannot say,
as a priori, whether our Universe will end with the big rip or
with the de Sitter phase, therefore it will be interesting to
study dynamical evolution of the Phantom models in general
potentials and the conditions for big rip and de Sitter phase
respectively.

\section{Big rip and de Sitter late time attractors}

Since current observations favor flat Universe, we will work in
the spatially flat Robertson-Walker metric. The corresponding
equations of motion and Einstein equations could be written as,

\begin{eqnarray}\label{sys}
&&\dot{H}=-\frac{\kappa^2}{2}(\rho_\gamma+p_\gamma-\dot{\phi}^2)\nonumber\\
&&\dot{\rho_\gamma}=-3H(\rho_\gamma+p_\gamma)\nonumber\\
&&\ddot{\phi}+3H\dot{\phi}-V'(\phi)=0\nonumber\\
&&H^2=\frac{\kappa^2}{3}(\rho_\gamma+\rho_{\phi})
\end{eqnarray}

\noindent where $\kappa^2=8\pi G$, $\rho_{\gamma}$ is the density
of fluid with a baryotropic equation of state
$p_{\gamma}=(\gamma-1)\rho_{\gamma}$, where $0\leq \gamma\leq2$ is
a constant that relates to the equation of state by $w=\gamma-1$;
The over dot represents derivative with respect to $t$, the prime
denotes derivative with respect to $\phi$.
$\rho_{\phi}=-\frac{1}{2}\dot{\phi}^{2}+V(\phi)$ and
$p_{\phi}=-\frac{1}{2}\dot{\phi}^{2}-V(\phi)$ are the energy
density and pressure of the $\phi$ field respectively, and $H$ is
Hubble parameter.

However, for an arbitrary potential, and considering the presence
of other energy density, one cannot find analytically solvable
models. Thus, we need to analyze the models via phase space
analysis. Similar as in Ref.\cite{attractor1,attractor2,macorra},
we introduce the following dimensionless variables
$x=\frac{\kappa}{\sqrt{6}H}\dot{\phi}$,
$y=\frac{\kappa\sqrt{V(\phi)}}{\sqrt{3}H}$,
$\lambda=-\frac{V'(\phi)}{\kappa V(\phi)}$,
$\Gamma=\frac{V(\phi)V''(\phi)}{V'^2(\phi)}$ and $N=\log a$. Then,
the equation system(\ref{sys}) could be reexpressed as the
following system of equations:

\begin{eqnarray}\label{auto}
\frac{dx}{dN}&=&\frac{3}{2}x[\gamma(1+x^2-y^2)-2x^2]-(3x+\sqrt{\frac{3}{2}}\lambda
y^2)\nonumber\\\frac{dy}{dN}&=&\frac{3}{2}y[\gamma(1+x^2-y^2)-2x^2]-\sqrt{\frac{3}{2}}\lambda
xy\nonumber\\\frac{d\lambda}{dN}&=&-\sqrt{6}\lambda^2x(\Gamma-1)
\end{eqnarray}

\noindent Also, we have a constraint equation

\begin{equation}\label{constraint}
\Omega_{\phi}+\frac{\kappa^2\rho_\gamma}{3H^2}=y^2-x^2+\frac{\kappa^2\rho_\gamma}{3H^2}=1
\end{equation}

The equation of state for the scalar fields could be expressed in
term of the new variables as

\begin{equation}\label{equaofstate}
 w_{\phi}=\frac{p_{\phi}}{\rho_{\phi}}=\frac{x^2+y^2}{x^2-y^2}
\end{equation}

Different from the case in exponential potential \cite{hao1}, the
parameters $\lambda$ and $\Gamma$ here are variables dependent on
$\phi$. Thus, strictly speaking, the above system is not an
autonomous system. So, just as was done in Refs.
\cite{attractor1,attractor2}, if we want to discuss the phase
plane, we need certain constraints on the potential, or
equivalently the condition under which the potential has the
property we want, in order that we can get some explicit results.

In the following, we will read out the big rip attractor as well
as de Sitter attractor from Eqs.(\ref{auto}) and specify the
corresponding conditions for the potential. The big rip attractor
is the dynamical attractor that corresponds to the Phantom
domination $\Omega_{\phi}=1$ and an equation of state $w<-1$.
While, the de Sitter attractor also corresponds to Phantom
domination but with $w=-1$. Looking at Eqs.(\ref{auto}), one can
observe that physically meaningful critical points $(x_c, y_c,
\lambda_c)$ of the system are (Note that here we are only
interested in the expansionist solutions ($y>0$))

\noindent (i). $(-\frac{\lambda_c}{\sqrt{6}},
\sqrt{1+\frac{\lambda_c^2}{6}}, \lambda_c)$ with $\Gamma=1$ and
$\lambda_c\neq 0$

\noindent (ii). $(0, 1, 0)$ for all $\Gamma$.

\noindent (iii). $(0, 0, \lambda_c)$ for all $\Gamma$.

\noindent where $\lambda_c$ could be fixed by $\Gamma=1$. To gain
some insight into the property of the critical points, we write
the variables near the critical points $(x_c, y_c, \lambda_c)$ in
the form $x=x_c+u$, $y=y_c+v$, and $\lambda=\lambda_c+\delta$ with
$u,v,\delta$ the perturbations of the variables near the critical
points. Substitute the expression into the system of equations
(\ref{auto}), one can obtain the equations for the perturbations.
The coefficients of the perturbation equations form a $3\times 3$
matrix $M$ whose eigenvalues determine the type and stability of
the critical points.

(i) Big rip attractor: in the above case (i), when $\Gamma\simeq
1$, from Eq.(\ref{auto}), we know that $\lambda$ is almost
constant. So, similar as was done in Refs.
\cite{attractor1,attractor2}, we consider only the first two
equations of Eqs.(\ref{auto}). Thus, the corresponding matrix $M$
should be two dimensional and the corresponding eigenvalues are
$(-3-\frac{\lambda_c^2}{2}, -3\gamma-\lambda_c^2)$, which indicate
that this critical point is a dynamical attractor of the system.
It is not difficult to evaluate that at the critical point, the
cosmic energy density parameter $\Omega_{\phi}=1$ and the equation
of state $w=-1-\frac{\lambda_c^2}{3}$. So, this attractor
corresponds to an equation of state $w<-1$ and will lead to the
catastrophic big rip. It is worth noting that for an arbitrary
potential, $\lambda_c$ is not a constant, which could be
determined by solving $\Gamma=1$ for a given potential. While in
the exponential potential, $\lambda_c$ is denoted as $\lambda$ and
is a constant parameter of the potential. For a general potential
$V(\phi)$, the condition for the existence of the above mentioned
big rip attractor is $\Gamma= 1$(for practical purpose, this
condition could be $\Gamma\simeq 1$, as was done in Refs.
\cite{attractor1,attractor2}) and $\lambda\neq 0$.

(ii) de Sitter attractor: in the above case (ii), the
corresponding eigenvalues of matrix $M$ are $(-3, 0, -3\gamma)$,
which indicate that the critical point is a dynamical attractor
corresponding to equation of state $w=-1$ and cosmic energy
density parameter $\Omega_{\phi}=1$, which is a de Sitter
attractor\cite{hao2}. The condition of such a de Sitter attractor
is $\lambda=0$. According to the previous definition for
$\lambda$, this condition is equivalent to that the attractor
appears at the extremum of the potential. We must point out that
when keeping the perturbation to the linear order, the right hand
side of the third equation in Eqs.(\ref{auto}) will vanish. Thus
the condition $\lambda=0$ is only a less restrictive one, from
which we cannot determine if the extremum is maximum or minimum.
To extract more information, one needs to either keep higher order
perturbation or choose different dimensionless variables as was
done in Ref. \cite{hao2}, in which we choose
$x=\frac{\phi}{\phi_0}$, $y=\frac{\dot{\phi}}{\phi_0^2}$ and
reduce Eqs.(\ref{sys}) to a different form. The analysis indicates
that the extremum must be a maximum for the existence of de Sitter
attractor for Phantom models. For details, please refer to Ref.
\cite{hao2}.

In the above case (iii), the critical point corresponds to
$\Omega_{\phi}=0$ and the corresponding definition of $w_{\phi}$
becomes meaningless. Moreover, the corresponding eigenvalues of
the $M$ matrix is $(0, 3(\gamma-2)/2, 3\gamma/2)$, which indicates
the critical point is not a dynamical attractor. So, we won't
address it into details.

Next, let's consider two specific models that admit big rip
attractor and de Sitter attractor respectively. To do so, we must
specify the potential. For the big rip attractor, we have the
requirement that $\lambda\neq 0$ and $\Gamma=1$ (or $\Gamma\simeq
1$ as approximation for practical purpose). It won't be difficult
to find that the solution for the exact $\Gamma=1$ would be that
$V(\phi)=C_1e^{C_2 \phi}$ with $C_1$ and $C_2$ as arbitrary
constants. Conventional study for quintessence models focus on the
case that $C_2<0$. Here, $C_2$ could be both positive and negative
constant. For definite, we choose $V(\phi)=V_0e^{\alpha\phi}$ with
$\alpha$ a positive constant, and thus
$\lambda=-\frac{\alpha}{\kappa}$. In Fig.1 and Fig.2, we plot the
numerical results.
\begin{figure}
\epsfig{file=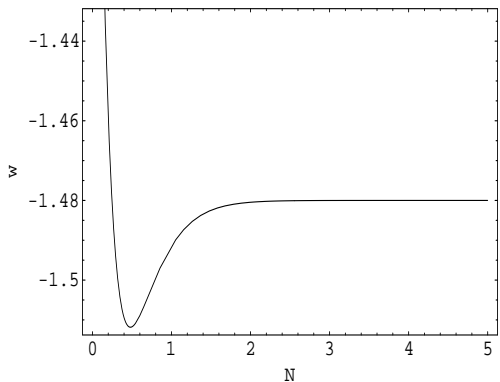,height=2.0in,width=2.6in} \caption{The
evolution of equation of state $w$ vs. $N$. We choose
dimensionless parameter $\frac{\alpha}{\kappa}=1.2$ and the
$\gamma=1$.}
\end{figure}

\begin{figure}
\epsfig{file=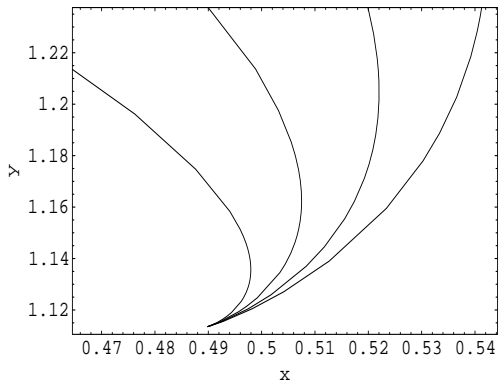,height=2.0in,width=2.6in} \caption{Phase
graph of the model for different initial $x$ and $y$. We choose
dimensionless parameter $\frac{\alpha}{\kappa}=1.2$ and the
$\gamma=1$. }
\end{figure}

For the de Sitter attractor, we require $\lambda=0$, which is
equivalent to that the critical point is the extremum of the
potential. We choose the potential
\begin{equation}\label{potentialha1}
V(\phi)=V_0-\sigma_0(\frac{\phi}{\phi_0})^2
\end{equation}
\noindent as a toy model, in which the de Sitter attractor
corresponds to the maximum of the potential
Eq.(\ref{potentialha1}). The corresponding numerical results
listed in Fig.3 and Fig.4
\begin{figure}
\epsfig{file=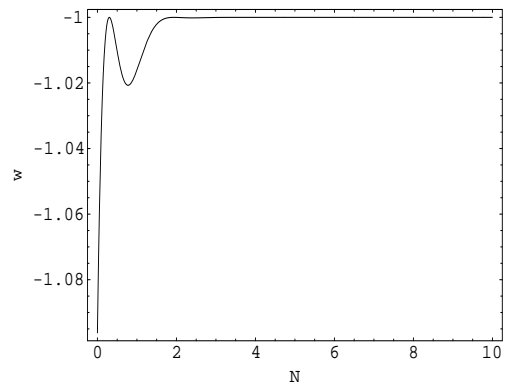,height=2.0in,width=2.6in} \caption{The
evolution of equation of state $w$ vs. $N$, in which for
simplicity, we choose $V_0=\phi_0=\kappa=\sigma_0=\gamma=1$}
\end{figure}

\begin{figure}
\epsfig{file=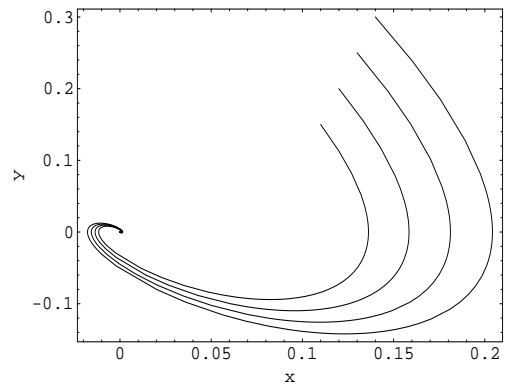,height=2.0in,width=2.6in} \caption{The phase
graph of the system with $x=\phi/\phi_0$ and
$y=\dot{\phi}/\phi_0^2$, which are so defined for convenience of
evaluation. The condition for evaluation is
$V_0=\phi_0=\kappa=\sigma_0=\gamma=1$.}
\end{figure}

\section{Tracking attractor}
In this section, we will show the tracking attractor behavior of
the dynamical system that are not manifested in the above
analysis. We consider the situation that $\lambda$ is very large
and $\Gamma$ is not 1 but nearly constant. It will be convenient
for the discussion if we make the following transformation
$\epsilon=\frac{1}{\lambda}$, $x=\epsilon X$ and $y=\epsilon
Y$\cite{attractor2}. Then the dynamical system could be rewritten
in terms of the new variables $X$ and $Y$ as:
\begin{eqnarray}\label{newauto}
\frac{dX}{dN}&&=-\sqrt{6}(\Gamma-1)X^2+\frac{3}{2}\gamma
X-3X-\sqrt{\frac{3}{2}}Y^2\\\nonumber
\frac{dY}{dN}&&=-\sqrt{6}(\Gamma-1)XY-\sqrt{\frac{3}{2}}XY+\frac{3}{2}\gamma
Y
\end{eqnarray}

\noindent where we have omitted the terms contain $\epsilon$. The
critical points of the above autonomous system Eq.(\ref{newauto})
corresponding to the tracking solution is $(X_c,
Y_c)=(\sqrt{\frac{3}{2}}\gamma_e,
\sqrt{\frac{3}{2}\gamma_e(\gamma_e-2)})$, where $\gamma_e\equiv
\frac{\gamma}{2\Gamma-1}$ is a defined effective barotropic index
of the scalar field. The corresponding cosmic energy density
parameter for phantom is
$\Omega_{\phi}=-\frac{3\gamma_e}{\lambda^2}$ and the equation of
state for phantom is $w_{\phi}=w_e$ with $w_e\equiv \gamma_e-1$.
Clearly, to make $\Omega_{\phi}$ physically meaningful, we need
$\Gamma<1/2$ and therefore $\gamma_e<0$. Since $\lambda$ is large,
then $\Omega_{\phi}$ is small and as a consequence, the background
dominates. Also, at this point, the energy density parameter of
the phantom field will track the effective barotropic index of the
scalar field $\gamma_e$, by which the name of tracking solution is
justified. It is not difficult to observe that although the
equation of state of the background cosmological fluid is positive
(0 for matter and 1/3 for radiation), the effective equation of
state could be less than -1, which is required for the the cosmic
density parameter for phantom $\Omega_{\phi}$ to make sense. This
tracking property is very different from that of quintessence
models in which the corresponding critical point is $(X_c,
Y_c)=(\sqrt{\frac{3}{2}}\gamma_e,
\sqrt{\frac{3}{2}\gamma_e(2-\gamma_e)})$ with the $\gamma_e$
defined the same as that in the above phantom case. But the
corresponding cosmic density parameter is
$\Omega_{quint}=\frac{3\gamma_e}{\lambda^2}$ which is just the
opposite of the expression for phantom. This sign difference lead
to different requirements for $\Gamma$, that is, for quintessence,
$\Gamma$ must be greater than $1/2$ while for phantom $\Gamma$
must be less than $1/2$ so as to make their respective cosmic
density parameters physically meaningful. This constraint on the
potentials for phantom excludes the exponential
potential($V(\phi)=V_0e^{\alpha\phi}$ with $\Gamma=1$)and inverse
power law potential($V(\phi)=A/\phi^{n}$ with
$\Gamma=1+\frac{1}{n}$) that have been used in quintessence
models. A simple potential that satisfies the above condition
takes the form $V(\phi)=A\phi^{1+\frac{1}{n}}$ with $n>1$. In this
case, $\Gamma=\frac{1}{1+n}$.

Next, let's show the stability of the critical point. To do so, we
use the similar approach used in previous section and find out the
eigenvalues of the perturbation equation of the dynamical system
near the critical point. We write the $X=X_c+U$ and $Y=Y_c+V$,
with $U$, $V$ the perturbations. Then the equation system could be
written as, up to the first order of the perturbations,
\begin{eqnarray}\label{newperturb}
\frac{dU}{dN}&&=[-3+(9/2-3\Gamma)\gamma_e]U-3\sqrt{\gamma_e(\gamma_e-2)}V\\\nonumber
\frac{dV}{dN}&&=-3/2(2\Gamma-1)\sqrt{\gamma_e(\gamma_e-2)}U
\end{eqnarray}

\noindent The eigenvalues are
\begin{eqnarray}\label{newperturb}
\lambda_1&&=-\frac{3}{4}\bigg[2+(2\Gamma-3)\gamma_e\\\nonumber &&+
\sqrt{4+(4-24\Gamma)\gamma_e+(1+2\Gamma)^2\gamma_e^2}\bigg]\\\nonumber
\lambda_2&&=-\frac{3}{4}\bigg[2+(2\Gamma-3)\gamma_e\\\nonumber&&-
\sqrt{4+(4-24\Gamma)\gamma_e+(1+2\Gamma)^2\gamma_e^2}\bigg]
\end{eqnarray}

\noindent According to our previous discussion, we know that
$\Gamma<1/2$ is required for the solution physically meaningful.
Closely examine the two eigenvalues, one can find that the first
one always has a negative real part. While for the second one, the
term in the square root become negative if
$\Gamma<\frac{5-4\sqrt{2}}{14}$when $\gamma=1$(matter dominated
epoch) and $\Gamma<\frac{19-8\sqrt{6}}{46}$ when $\gamma=4/3$
(radiation dominated epoch). In these two cases, the critical
point is stable. While for the range of
$\frac{5-4\sqrt{2}}{14}\leq\Gamma<1/2$ and
$\frac{19-8\sqrt{6}}{46}\leq\Gamma<1/2$ corresponding to the
matter and radiation dominated epoch respectively, we plot the
second eigenvalue vs. $\Gamma$ in Fig.5 and Fig. 6. From the
plots, one can see that both cases yield negative eigenvalues and
therefore the critical point is stable for the range $\Gamma<1/2$.
That is, the tracking solution is a dynamical attractor of the
system.
\begin{figure}
\epsfig{file=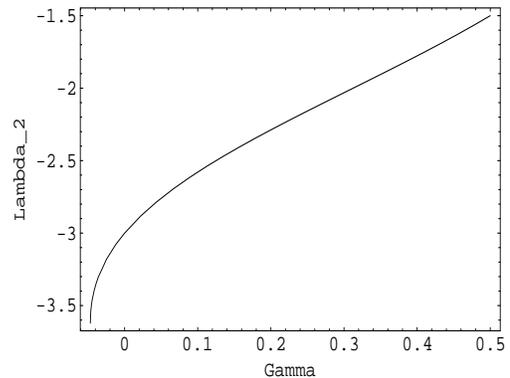,height=2.0in,width=2.6in} \caption{The
second eigenvalue $\lambda_2$ vs. $\Gamma$ at the matter dominated
epoch.}
\end{figure}
\begin{figure}
\epsfig{file=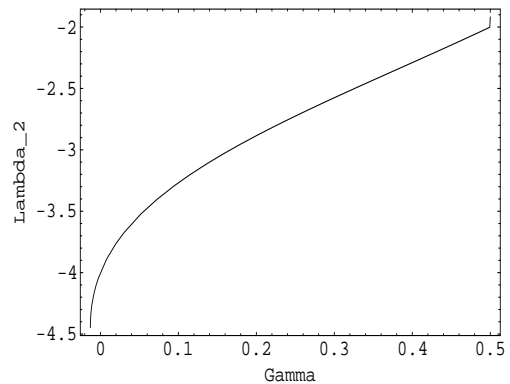,height=2.0in,width=2.6in} \caption{The
second eigenvalue $\lambda_2$ vs. $\Gamma$ at the radiation
dominated epoch.}
\end{figure}

\section{tachyonic Stability near the attractors}

In this section, we will study the "tachyonic"
instability\footnote{We call the system is tachyonically instable
if the effective mass of the perturbation becomes imaginary} of
the dynamical system near the dynamical attractors. The reason for
us to discuss the stability only near the attractors is that we
consider that our universe is just at or near one of the
attractors. The stability of the system at the attractors does not
necessarily guarantee the stability during its evolution, but if
we took a anthropic philosophy and the above assumption, we can
say that we are just living in a universe where the phantom
evolves safely to its stable attractors. In synchronous gauge, the
perturbed equation of motion reads \cite{caldwell,carroll}

\begin{equation}\label{perturbationeq}
\delta \ddot{\phi_k}+3H\delta
\dot{\phi_k}+(k^2-V''(\phi))=-\frac{1}{2}\dot{h}\dot{\phi}
\end{equation}

\noindent where the $h$ is the trace of $h_{ij}$ (the metric
perturbation) and $\delta \phi_k$ is a fourier mode of the phantom
field perturbation. The prime denotes the derivative with respect
to $\phi$ and the dot denotes the derivative with respect to $t$.
The effective mass for the perturbation is
$[k^2-V''(\phi)]^{1/2}$. For the potentials with negative
$V''(\phi)$, it will be always stable. While for potentials with
positive $V''$,the stability requires that the wave number of the
perturbation should be $k<k_{crit}=\sqrt{V''(\phi)}$. In terms of
the dimensionless variables we introduced previously, the critical
wave number could be written as

\begin{equation}\label{crick}
k_{crit}=\sqrt{V''(\phi)}=\sqrt{3}Hy|\lambda|\sqrt{\Gamma}
\end{equation}

\noindent At the tracking attractor, we can easily obtain that
$y_c=\sqrt{\frac{3\gamma_e(\gamma_e-2)}{2\lambda_c^2}}$ and the
corresponding critical wave number will be
$k_{crit}=3H\sqrt{\frac{\Gamma\gamma_e(\gamma_e-2)}{2}}$. If we
choose the Hubble parameter $H$ as the present value and
$\gamma_e$ as well as $\Gamma$ are of order 1, then the stability
will appear if the wave length of the perturbation is greater than
about $10^{28}$ cm, which is even greater than the radius of our
observable Universe. That is to say, the tracking attractor
solution corresponds to a state that is "tachyonically" stable. At
the big rip attractor, the critical wave number should be
$k_{crit}=|\lambda_c| H\sqrt{3+\frac{\lambda_c^2}{2}}$, with
$\lambda_c$ a dimensionless variable that can be specified by
solving $\Gamma=1$. If we also choose the Hubble parameter $H$ as
the present value and $\lambda_c$ is of order 1, then the
situation will be similar to the above case for tracking attractor
and the model is stable for practical purpose at the big rip
attractor. At the de Sitter attractor, we have
$V''(\phi_{crit})<0$ so that the point corresponds to the maximum
of the potential \cite{hao2}. Therefore, the models will always be
stable at the de Sitter attractor. From above discussion, we know
that the de Sitter phase, big rip phase and tracking phase all
could not only be dynamical attractors under properly constrained
potentials but also stable state in terms of the tachyonic
stability. However, as we have mentioned at the beginning of this
section, the stability at the attractors does not guarantee its
evolution. But this could be circumvented by adopting an anthropic
view.

\section{quantum stability}

Besides the tachyonic instability we have analyzed in section IV,
there is also another instability, quantum instability, that
plagued phantom models\cite{carroll}. Since dark energy is assumed
to be weakly interacted with other matter except via gravity, it
would be natural to consider only the phantom decay relevant to
gravitons, i.e. a phantom decays into other phantoms and
gravitons. Clearly, the simplest process will be one phantom
decays into two other phantoms and a graviton as

\begin{equation}\label{decay}
\phi_i\rightarrow h+\phi_1+\phi_2
\end{equation}

\noindent Follow the discussions in Ref. \cite{carroll}, we
firstly consider the decay in the tree level with potential
coupling and calculate the corresponding $\lambda_{eff}$ for the
potentials in this paper, i.e. exponential potential
$V(\phi)=V_0e^{\alpha\phi}$, quadratic potential Eq
.(\ref{potentialha1}) and an analytically solvable potential(its
analysis will be presented in section VI).
\begin{equation}\label{depot}
V(\phi)=V_0\{\sin[\sqrt{3}\kappa(\phi-\phi_0)]+\csc[\sqrt{3}\kappa(\phi-\phi_0)]\}
\end{equation}

It is not difficult to evaluate that the corresponding
$\lambda_{eff}$ are

\begin{equation}\label{lambdaeff}
\lambda_{eff}=
\left\{%
\begin{array}{ll}
    \frac{\alpha^3V(\phi_0)}{6M_p}, & \hbox{for exponential potential} \\
    0 \hspace{1cm}, & \hbox{for quadratic potential} \\
    \frac{50V_0\kappa^3}{40\sqrt{3}M_p} , & \hbox{for potential Eq.(\ref{depot})} \\
\end{array}%
\right.
\end{equation}

\noindent Similar as in Ref.\cite{carroll}, $\phi_0$ is of the
order of $M_p$. Note that the parameters in the potentials are of
the order $\alpha\sim\kappa\sim 1/M_p$, the $\lambda_{eff}$ for
exponential potential as well as the solvable potential
(\ref{depot}) are both of the order $V_0/M_p^4\sim 10^{-120}$, the
same as that in Ref.\cite{carroll} and thus need the same momentum
cutoff. But for the quadratic potential, $\lambda_{eff}=0$ and
therefore the corresponding model need no momentum cutoff and is
stable in the tree level potential coupling. As we have clarified
at the beginning of this section, the decay described above is the
simplest one and more complicated decay process will need higher
derivative of the potentials. So, the quadratic potential model is
stable if we consider only the tree level potential coupling. But
the other two potentials need momentum cutoff of
$\Lambda<10^{60}M_p$, which is still acceptable\cite{carroll}.

While, if we also consider the derivative coupling, which is
potential independent, the situation will be the same as that in
Ref.\cite{carroll} and the optimistic estimation of the momentum
cutoff for the models should be less than $10^{20}M_p\sim 100$MeV.
However, this is not necessarily impossible for phantom if we
assume the scalar field theory is only valid up to relatively low
momenta or find some mechanism to suppress derivative couplings,
leaving only the couplings of gravitons to the potential, which
were consistent with a cutoff as high as the Planck
scale\cite{carroll}. The point is that if the observations
eventually do suggest dark energy with an equation of state $w<-1$
and if it is modelled by phantom, then it must be stable with some
mechanisms such as those described above.

\section{Two analytical models}

In the following, to expose the above discussion more
specifically, we discuss two solvable Phantom models that
correspond to quasi-de Sitter and big rip late time phase
respectively. We consider only the case that Phantom become
dominant. The potential for the model corresponding to quasi-de
Sitter solution is in the periodic potential Eq.(\ref{depot}). For
simplicity, we consider only the behavior of $V(\phi)$ for
$\phi\in[\phi_0, \phi_0+\frac{\pi}{\sqrt{3}\kappa}]$. The quasi-de
Sitter solutions are
\begin{eqnarray}\label{solu1}
a&=&a_0\{\sec[\sqrt{3}\kappa(\phi-\phi_0)]\}^{1/3}\\\nonumber t&=&
t_0+\frac{1}{(288V_0^2)^{\frac{1}{4}}\kappa}\{\ln\bigg(\frac{1+[1-(\frac{a_0}{a})^6]
^{\frac{1}{4}}}{1-[1-(\frac{a_0}{a})^6]^{\frac{1}{4}}}\bigg)\\\nonumber&&+2\arctan[1-(\frac{a_0}{a})^6]^{\frac{1}{4}}\}\\\nonumber
\rho_{\phi}&=&\sqrt{2}V_0[1-(\frac{a_0}{a})^6]\\\nonumber
w&=&-\frac{1}{1-(a_0/a)^6}
\end{eqnarray}

\noindent It is easily found that the scale factor $a(t)$ tends to
infinite as $\phi\rightarrow\phi_0+\frac{\pi}{2\sqrt{3}\kappa}$.
Furthermore, as the Universe expands, the Phantom energy density
$\rho_{\phi}$ tends to $\sqrt{2}V_0$ and the equation of state
$w_{\phi}$ approaches $-1$.

The second potential for the model admitting big rip solution is
\begin{equation}\label{bigpot}
V(\phi)=V_0\exp[\sqrt{3}\kappa A(\phi-\phi_0)]
\end{equation}
\noindent where $V_0$ and $A$ are positive constant. The solutions
of the system are

\begin{eqnarray}\label{solu2}
a&=&a_0\exp[\frac{\kappa}{\sqrt{3}A}(\phi-\phi_0)]\\\nonumber
t&=&t_0+\frac{2(2+A^2)^{1/2}}{\sqrt{3}\kappa(2V_0)^{1/2}A^2}[1-(\frac{a_0}{a})^{3A^2/2}]\\\nonumber
\rho_{\phi}&=&\frac{2V_0}{2+A^2}(\frac{a}{a_0})^{3A^2}\equiv
\rho_0(\frac{a}{a_0})^{3A^2}\\\nonumber w_{\phi}&=&-(1+A^2)
\end{eqnarray}

\noindent Using Eq.(\ref{solu2}), we find that the remaining time
$t_r=t_{end}-t_0$ before the end of the Universe is given
analytically by
\begin{equation}\label{tr}
t_r=\frac{2}{\sqrt{3\rho_0}\kappa A^2}
\end{equation}

\noindent In Eq.(\ref{tr}), putting in
$\kappa\sqrt{\rho_0}=\sqrt{3}\Omega_{\phi}H_0$ with
$\Omega_{\phi}=0.73$ and $H_0^{-1}=13.7Gyr$, one finds
$t_r=28.01$, 53.4 and 106.9 Gyr. for $w=-1.38$, -1.20 and -1.10
respectively.

\section{Discussion}
We study the phase space of phantom model in an arbitrary
potential and give the corresponding conditions for tracking
attractor, big rip attractor and de Sitter attractor. The big rip
attractor and de Sitter attractor are two late time attractors
that will determine the type of the cosmic doomsday. For the de
Sitter attractor, we have $\lambda=0$ and therefore a divergent
$\Gamma$ according to its definition. However, in the equations of
motion Eqs.(\ref{auto}), such singularity will not appear because
of the term $\lambda^2\Gamma=\frac{V''}{\kappa^2}$ gives regular
value. On the other hand, the tracking attractor serves a similar
function as it does in traditional quintessence models and could
alleviate the fine tuning of the model. The observational
constraint $-1.38<w<0.82$ is often interpreted as an observational
support for the existence of Phantom in our Universe. We show that
there maybe two possible phantom cosmic doomsday, big rip and de
Sitter, which are both attractor phases of the dynamical system
and compatible with current observations. At least before the
$w<-1$ dark energy is excluded by observation completely, the
phantom models will remain an interesting alternative.

\vspace{0.8cm} \noindent ACKNOWLEDGEMENT: This work is supported
by NKBRSF under Grant No. 1999075406.

\end{document}